\documentstyle[11pt]{article}

\newcommand{\EQ}{\begin{equation}}
\newcommand{\EN}{\end{equation}}
\newcommand{\bea}{\begin{eqnarray}}
\newcommand{\ena}{\end{eqnarray}}
\newcommand{\vs}[1]{\vspace{#1 mm}}
\newcommand{\hs}[1]{\hspace{#1 mm}}

\newcommand{\vt}{\vartheta}
\newcommand{\vp}{\varphi}
\def\bbox{{\,\lower0.9pt\vbox{\hrule \hbox{\vrule height 0.2 cm
\hskip 0.2 cm \vrule height 0.2 cm}\hrule}\,}}
\newcommand{\dsl}{\pa \kern-0.5em /}

\newcommand{\pa}{\partial}

\newcommand{\nn}{\nonumber\\}
\newcommand{\p}[1]{(\ref{#1})}

\begin{document}

\topmargin 0pt
\oddsidemargin 5mm

\renewcommand{\thefootnote}{\fnsymbol{footnote}}
\begin{titlepage}

\setcounter{page}{0}
\begin{flushright}
OU-HET 279 \\
UB-ECM-PF-97/30\\
hep-th/9710129
\end{flushright}

\vs{10}
\begin{center}
{\Large SUPERSYMMETRY OF M-BRANES AT ANGLES}
\vs{15}

{\large
Nobuyoshi Ohta\footnote{e-mail address: ohta@phys.wani.osaka-u.ac.jp}
} \\
\vs{5}
{\em Department of Physics, Osaka University, \\
Toyonaka, Osaka 560, Japan} \\
\vs{5}
and \\
\vs{5}
{\large
Paul K. Townsend\footnote{e-mail address: pkt@ecm.ub.es. On leave from DAMTP,
University of Cambridge, U.K.}
} \\
\vs{5}
{\em Departament ECM, Facultad de F\'{\i}sica,\\
 Universitat de Barcelona and Institut de F\'{\i}sica dAltes Energies, \\
Diagonal 647, E-08028 Barcelona, Spain }\\ 
\end{center}
\vs{10}
\centerline{{\bf{Abstract}}}

We determine the possible fractions of supersymmetry preserved by
two intersecting M-5-branes. These include the fractions 3/32 and 5/32 which
have not occurred previously in intersecting brane configurations. Both occur
in non-orthogonal pointlike intersections of M-5-branes but $5/32$
supersymmetry is possible only for specific fixed angles.

\end{titlepage}
\newpage
\renewcommand{\thefootnote}{\arabic{footnote}}
\setcounter{footnote}{0} 

\section{Introduction}

It has become clear over the past few years that many supersymmetric quantum
field theories may be realized as worldvolume theories on branes, or on their
intersections.  One is tempted to conjecture that all anomaly-free
interacting supersymmetric quantum field theories (without gravity) may be
realized in this way. Although we shall not attempt to establish this
conjecture here, it provides the motivation for the work that we shall
report on. Interacting supersymmetric field theories without gravity are
restricted to spacetime dimensions $D\le10$, and the number of
supersymmetries for each value of $D$ is also severely restricted. In the
context of branes it is convenient to refer to this number as a fraction
$\nu$ of the supersymmetry of the M-theory vacuum, which is maximally
supersymmetric. For example, M-branes preserve 1/2 supersymmetry, as do
the D-branes of superstring theory and various other branes:
together, these may be considered as the `basic' branes of M-theory
and its superstring duals. The worldvolume field theories of these $\nu=1/2$
branes are either dimensional reductions of $D=10$ super Maxwell theory or,
in the case of the M-5-brane or type II NS-5-branes, the $D=6$
(2,0)-supersymmetric antisymmetric tensor theory. These are
non-interacting field theories but the interacting versions are found
as theories on coincident parallel branes\footnote{The $D=10$ super
Yang-Mills theory does, strictly speaking, not have a brane interpretation
but it is also anomalous.}. There are no other field theories with this
fraction of supersymmetry (where by `field theories' it should now be
understood that we mean theories without gravity).

There are of course plenty of other supersymmetric field theories with
$\nu<1/2$. Many of these are known to have an interpretation as worldvolume
field theories on the intersection of two or more branes. In the case of
orthogonal intersections the determination of the fraction of supersymmetry
preserved is straightforward: only the fractions $\nu=1/4,\, 1/8,\, 1/16,\, 
1/32$ occur. All these fractions are known to be realized by supersymmetric
field theories in various dimensions but various other fractions are also
possible when $D\le3$. For example, $\nu=3/16$ is realized in $D=3$ by
topologically-massive super Yang-Mills theory~\cite{ZK,KL}. This was
recently argued to be the worldvolume field theory on certain
non-orthogonal intersections of two IIB 5-branes preserving $3/16$
supersymmetry~\cite{GG}. As shown in~\cite{GG}, these configurations are
dual to certain non-orthogonal $D=2$ intersections of two M-5-branes,
for which the effective field theory is presumably the (3,3)-supersymmetric
dimensional reduction of the $D=3$ $N=3$ supersymmetric gauge theory.
The M(atrix) theory description of these $\nu=3/16$ configurations has
also been found recently~\cite{OZ}.

Various other values of $\nu$ can be realized by $D=2$ supersymmetric
sigma-models with $(p,q)$ supersymmetry. For example (1,2) supersymmetry
yields $\nu=3/32$ while (1,4) supersymmetry yields $\nu=5/32$. Since
$p,q=0,1,2,4$, the fractions $\nu=k/32$ with $k=1,2,3,4,5,6,8$ can be found
in this way\footnote{The fraction $\nu=3/16$ arises from the possibility
of (2,4) supersymmetry. It is not clear at present whether such field
theories can be realized as worldvolume field theories on intersecting
branes. If so, it must be that the branes are intrinsically intersecting,
as against merely overlapping, since if they could be pulled apart all
fields would be massive and the intersection field theory could not be
chiral.}. Note that the fraction $\nu=7/32$ is absent; in fact, there is
no known $\nu=7/32$ supersymmetric field theory. On the other hand,
the fractions $\nu=3/32$ and $\nu=5/32$, which do correspond to known $D=2$
(and hence also $D=1$) supersymmetric field theories, have not yet been
found as fractions of supersymmetry preserved by the intersection
of two branes. Clearly, they must be found if the conjecture that all
(anomaly-free) supersymmetric field theories are realizable as worldvolume
intersection field theories is to have a chance of being true. We shall
show here that these fractions are realized by a pair of M-5-branes
intersecting at certain angles. We accomplish this by an exhaustive
analysis of the fractions of supersymmetry preserved by a pair of
intersecting M-5-branes. Specifically, we shall show that the fraction
of supersymmetry preserved by such configurations necessarily takes
one of the following values:
\EQ
\nu= {1\over2},\ {1\over4},\ {3\over16},\ {5\over32},\ {1\over8},\
{3\over32},\ {1\over16},\ {1\over32}\, .
\label{fraction}
\EN
The fraction $1/2$ occurs only for parallel M-5-branes which, strictly
speaking, is not an `intersecting brane' configuration but it is
convenient to include this case as it will be the starting point for
obtaining all the other possibilities by rotation of one of the
M-5-branes. The fractions $5/32$, $3/32$ and $1/32$ occur only when the
tangent vectors to the two 5-branes span the entire ten-dimensional space.
Thus, these fractions cannot be realized by, for example, intersecting
M-2-branes. This is consistent with the fact that fractions of the form
$\nu= (2n+1)/32$ are possible only for pointlike ($D=1$) or stringlike
($D=2$) intersections; for intersecting M-5-branes the intersection is
pointlike. It is possible that these fractions are also realized by other
intersecting brane configurations with stringlike intersections, but it
is unlikely that fractions $\nu$ other than those listed above will be
found in this way. The current state of knowledge concerning the
conditions imposed by supersymmetry in various dimensions is
essentially complete\footnote{For example, a complete analysis for $(p,q)$
sigma-models can be found in~\cite{HP,PT1}. A partial analysis of $D=1$
sigma-models can be found in~\cite{CP1,GPS}; it would be desirable to have a
complete analysis for this case.}, and there is no known example with a number
of supersymmetries  other than those implied by \p{fraction}. Furthermore, it
seems likely that any configuration of two intersecting branes will be in the
same `duality equivalence class' as one involving only M-5-branes. This is
certainly true for orthogonal intersections for which there are just two
duality equivalence classes; the M-5-brane representatives are the
intersection of M-5-branes over a 3-plane~\cite{PT2, TS1} and M-5-branes
intersecting on a line~\cite{GKT}.

\section{M-5-branes at angles}

We start from two parallel M-5-branes. We can choose cartesian coordinates
such that these M-5-branes lie in the 12349 5-plane. This configuration is
summarised by the array
\EQ
\begin{array}{lcccccccccc}
M: & 1 & 2 & 3 & 4 & - & - & - & - & 9 & - \\
M: & 1 & 2 & 3 & 4 & - & - & - & - & 9 & - 
\end{array}\, ,
\label{0}
\EN
and is associated with the constraint
\bea
\label{1}
\Gamma_{091234}\epsilon = \epsilon\, .
\ena
The $SO(1,10)$ spinors $\epsilon$ satisfying this relation can be considered
as the asymptotic values of Killing spinor fields of an associated $D=11$
supergravity solution. We shall therefore refer to them as `Killing spinors'.
The parallel M-5-branes may be coincident or they may be separated by some
distance in the transverse directions. The distinction will not be of
relevance to the following discussion, but if the M-5-branes are not
coincident the rotation of one will lead to a configuration of `overlapping',
rather than intersecting, branes. For convenience we ignore this distinction
here, and refer only to `intersecting' branes. 

We now fix one M-5-brane, the `first', and rotate the second one. Denoting the
spinor representation of the rotation matrix for the second M-5-brane by
$R$, we now have an additional constraint
\cite{BDL}
\bea
\label{2}
R\Gamma_{091234}R^{-1}\epsilon = \epsilon\, .
\ena
As explained in \cite{GG, PK}, $R$ effectively depends on five independent
angles and can be chosen to take the form
\bea
\label{3}
R = e^{{1\over2}[\vt \Gamma_{15} + \psi\Gamma_{26} + 
\vp\Gamma_{37} + \rho\Gamma_{48} + \zeta\Gamma_{9\natural}]}\, ,
\ena
where we use the symbol $\natural$ for the number 10, and $\vt,\psi,\vp,\rho$
and $\zeta$ are the five angles characterising the rotation. Note that
$\Gamma_{091234}R^{-1} = R\Gamma_{091234}$, so that \p{2} becomes
$R^2\Gamma_{091234}\epsilon=\epsilon$. Hence the condition (\ref{2}) becomes
\bea
\label{4}
[R^2 -1]\epsilon =0,
\ena
with $R$ given by \p{3}. 

We wish to determine the number of simultaneous solutions (equal to
$32\nu$) of \p{1} and \p{4} as a function of the five angles characterising
the rotation matrix $R$. The previously known partial solutions to this
problem were summarised in \cite{PK}. Here we present the general solution.
We first note that
\bea
R^2 -1 &=& 2 R \Gamma_{15} \left[
\sin\frac{\vt}{2} \cos\frac{\psi}{2} \cos\frac{\vp}{2} \cos\frac{\rho}{2}
\cos\frac{\zeta}{2}
- \Gamma_{1526} \cos\frac{\vt}{2} \sin\frac{\psi}{2} \cos\frac{\vp}{2}
 \cos\frac{\rho}{2} \cos\frac{\zeta}{2} \right. \nn
&& \hs{-6} - \Gamma_{1537} \cos\frac{\vt}{2} \cos\frac{\psi}{2}
 \sin\frac{\vp}{2} \cos\frac{\rho}{2} \cos\frac{\zeta}{2}
- \Gamma_{1548} \cos\frac{\vt}{2} \cos\frac{\psi}{2} \cos\frac{\vp}{2}
 \sin\frac{\rho}{2} \cos\frac{\zeta}{2} \nn
&& \hs{-6}- \Gamma_{159\natural} \cos\frac{\vt}{2} \cos\frac{\psi}{2}
 \cos\frac{\vp}{2} \cos\frac{\rho}{2} \sin\frac{\zeta}{2}
- \Gamma_{1537489\natural} \cos\frac{\vt}{2} \cos\frac{\psi}{2}
 \sin\frac{\vp}{2} \sin\frac{\rho}{2} \sin\frac{\zeta}{2} \nn
&& \hs{-6}- \Gamma_{1526489\natural} \cos\frac{\vt}{2} \sin\frac{\psi}{2}
 \cos\frac{\vp}{2} \sin\frac{\rho}{2} \sin\frac{\zeta}{2}
- \Gamma_{1526379\natural} \cos\frac{\vt}{2} \sin\frac{\psi}{2}
 \sin\frac{\vp}{2} \cos\frac{\rho}{2} \sin\frac{\zeta}{2} \nn
&& \hs{-6}- \Gamma_{15263748} \cos\frac{\vt}{2} \sin\frac{\psi}{2}
 \sin\frac{\vp}{2} \sin\frac{\rho}{2} \cos\frac{\zeta}{2}
+ \Gamma_{489\natural} \sin\frac{\vt}{2} \cos\frac{\psi}{2}
 \cos\frac{\vp}{2} \sin\frac{\rho}{2} \sin\frac{\zeta}{2} \nn
&& \hs{-6}+ \Gamma_{379\natural} \sin\frac{\vt}{2} \cos\frac{\psi}{2}
 \sin\frac{\vp}{2} \cos\frac{\rho}{2} \sin\frac{\zeta}{2}
+ \Gamma_{3748} \sin\frac{\vt}{2} \cos\frac{\psi}{2}
 \sin\frac{\vp}{2} \sin\frac{\rho}{2} \cos\frac{\zeta}{2} \nn
&& \hs{-6}+ \Gamma_{269\natural} \sin\frac{\vt}{2} \sin\frac{\psi}{2}
 \cos\frac{\vp}{2} \cos\frac{\rho}{2} \sin\frac{\zeta}{2}
+ \Gamma_{2648} \sin\frac{\vt}{2} \sin\frac{\psi}{2}
 \cos\frac{\vp}{2} \sin\frac{\rho}{2} \cos\frac{\zeta}{2} \nn
&& \hs{-6}\left.+ \Gamma_{2637} \sin\frac{\vt}{2} \sin\frac{\psi}{2}
 \sin\frac{\vp}{2} \cos\frac{\rho}{2} \cos\frac{\zeta}{2}
+ \Gamma_{2637489\natural} \sin\frac{\vt}{2} \sin\frac{\psi}{2}
 \sin\frac{\vp}{2} \sin\frac{\rho}{2} \sin\frac{\zeta}{2}
\right].
\label{5}
\ena
Since the gamma matrix products appearing in \p{5} commute with each other
and with $\Gamma_{091234}$, we can simultaneously diagonalize all these
matrices, and since each of them squares to the identity their eigenvalues
are all $\pm 1$. Moreover, the traces of these matrices, and the traces of
products of pairs of them, vanish. We can therefore arrange for these
matrices to take the form
\bea
\Gamma_{09123} &=& {\rm diag.} (\overbrace{1,\cdots ,1}^{16},
 \overbrace{-1,\cdots, -1}^{16}), \nn
\Gamma_{1526} &=& {\rm diag.} (\overbrace{1,\cdots ,1}^{8},
 \overbrace{-1,\cdots, -1}^{8}, \cdots), \nn
\Gamma_{1537} &=& {\rm diag.} (\overbrace{1,\cdots ,1}^{4},
 \overbrace{-1,\cdots, -1}^{4}, \overbrace{1,\cdots ,1}^{4},
 \overbrace{-1,\cdots, -1}^{4}, \cdots), \nn
\Gamma_{1548} &=& {\rm diag.} (1,1,-1,-1,1,1,-1,-1,1,1,-1,-1,1,1,-1,-1,
 \cdots), \nn
\Gamma_{159\natural} &=& {\rm diag.} (1,-1,1,-1,1,-1,1,-1,1,-1,1,-1,1,-1,1,-1,
 \cdots),
\label{6}
\ena
the rest being determined by the products of those given.  Note that in this
basis the first condition \p{1} projects out the second 16 components of the
Killing spinor $\epsilon$, leaving just the first 16 components. Our task is
therefore to determine the consequences of the second condition \p{4} for the
first 16 components.

To proceed with the analysis we use \p{6} in \p{5} to derive the
following result:
\bea
R^2-1 &=& 2 R \Gamma_{15} \times {\rm diag.} \left(
 \sin\frac{\vt-\psi-\vp-\rho-\zeta}{2},
 \sin\frac{\vt-\psi-\vp-\rho+\zeta}{2},\right. \nn
&& \sin\frac{\vt-\psi-\vp+\rho-\zeta}{2},
 \sin\frac{\vt-\psi-\vp+\rho+\zeta}{2},
 \sin\frac{\vt-\psi+\vp-\rho-\zeta}{2},\nn
&& \sin\frac{\vt-\psi+\vp-\rho+\zeta}{2},
 \sin\frac{\vt-\psi+\vp+\rho-\zeta}{2},
 \sin\frac{\vt-\psi+\vp+\rho+\zeta}{2},\nn
&& \sin\frac{\vt+\psi-\vp-\rho-\zeta}{2},
 \sin\frac{\vt+\psi-\vp-\rho+\zeta}{2},
 \sin\frac{\vt+\psi-\vp+\rho-\zeta}{2},\nn
&& \sin\frac{\vt+\psi-\vp+\rho+\zeta}{2},
 \sin\frac{\vt+\psi+\vp-\rho-\zeta}{2},
 \sin\frac{\vt+\psi+\vp-\rho+\zeta}{2},\nn
&& \sin\frac{\vt+\psi+\vp+\rho-\zeta}{2},\left.
 \sin\frac{\vt+\psi+\vp+\rho+\zeta}{2}, \cdots \right),
\label{7}
\ena
where the last 16 components are omitted because, for the reason just given,
they are not needed for the determination of the fraction $\nu$ of unbroken
supersymmetry. We shall now use this result to provide a systematic analysis
of the possible values of $\nu$.

\subsection{One angle}

We begin with the simplest case of a rotation by single angle $\vt$.
Setting the other angles to zero, we have
\bea
R^2-1 &=& 2 R \Gamma_{15} \sin\frac{\vt}{2}\times {\rm diag.} (
{\bf 1}_{16}, \dots),
\label{8}
\ena
where ${\bf 1}_{16}$ is the $16 \times 16$ identity matrix. Supersymmetry
is completely broken unless $\sin\frac{\vt}{2}=0$, {\it i.e.} $\vt=0$
(mod $2\pi$).

\subsection{Two angles}

We now have
\bea
R^2-1 &=& 2 R \Gamma_{15} \times {\rm diag.} \left(
 \sin\frac{\vt-\psi}{2} \, {\bf 1}_8,\, \sin\frac{\vt+\psi}{2}\, {\bf 1}_8,
\cdots
\right),
\label{9}
\ena
where ${\bf 1}_{8}$ is the $8 \times 8$ identity matrix.
Supersymmetry is completely broken unless $\vt\pm \psi=0$ (mod $2\pi$).
When this condition is satisfied the fraction of unbroken supersymmetry is
$1/4$. This includes as a special case the orthogonal intersection over a
3-plane of two M-5-branes. We thus recover the result of \cite{BDL} that
rotations away from orthogonality can preserve $1/4$ supersymmetry.

\subsection{Three angles}

We now have
\bea
R^2-1 &=& 2 R \Gamma_{15} \times {\rm diag.} \left(
 \sin\frac{\vt-\psi-\vp}{2}\, {\bf 1}_4,\, \sin\frac{\vt-\psi+\vp}{2}\,
{\bf 1}_4, \right. \nn
&& \left. \sin\frac{\vt+\psi-\vp}{2}\, {\bf 1}_4,\,  
 \sin\frac{\vt+\psi+\vp}{2}\, {\bf 1}_4, \cdots \right),
\label{10}
\ena
where ${\bf 1}_4$ is the $4 \times 4$ identity matrix.
Supersymmetry is completely broken unless $\vt\pm \psi\pm \vp=0$ (mod $2\pi$).
When this condition is satisfied the fraction of unbroken supersymmetry is
$1/8$; an example involving 6-branes was given in \cite{BDL}. Note that
the condition on the three angles imposed by supersymmetry does not allow
a rotation to a configuration of orthogonally intersecting branes. This
shows that rotations away from orthogonality do not yield all possible
supersymmetric configurations of branes intersecting at angles; one must
instead consider rotations away from parallel branes, as we are doing here.

\subsection{Four angles}

We have
\bea
R^2-1 &=& 2 R \Gamma_{15} \times {\rm diag.} \left(
 \sin\frac{\vt-\psi-\vp-\rho}{2}\, {\bf 1}_2,\, 
\sin\frac{\vt-\psi-\vp+\rho}{2}\, {\bf 1}_2,\right. \nn
&& \hs{-6}\sin\frac{\vt-\psi+\vp-\rho}{2}\, {\bf 1}_2,\,
 \sin\frac{\vt-\psi+\vp+\rho}{2}\, {\bf 1}_2,\,
\sin\frac{\vt+\psi-\vp-\rho}{2}\, {\bf 1}_2, \nn
&& \hs{-6}\sin\frac{\vt+\psi-\vp+\rho}{2}\, {\bf 1}_2,\,
\sin\frac{\vt+\psi+\vp-\rho}{2}\, {\bf 1}_2,\,
 \left. \sin\frac{\vt+\psi+\vp+\rho}{2}\, {\bf 1}_2, \cdots \right),
\label{11}
\ena
where ${\bf 1}_2$ is the $2 \times 2$ identity matrix.

Supersymmetry is completely broken unless $\vt\pm \psi\pm \vp\pm \rho=0$
(mod $2\pi$). We are free to change the signs of the angles, so without loss
of generality we may set
\EQ
\rho =\vt -\psi+\vp\, .
\EN
The generic configuration of this type preserves $1/16$ supersymmetry,
a fraction found previously only in orthogonal intersections of four
branes. Thus, these are new supersymmetric intersecting brane configurations.
In special cases the supersymmetry is enhanced. For example we have
(generically) $1/8$ supersymmetry for the following special values
considered in~\cite{PK}:
\EQ
\cases{ & $\vt = \psi \neq \vp =\rho$ \, , \cr
{\rm or} \ & $\vt= \rho \neq \psi = \vp$ \, , \cr
{\rm or} \ & $\vt= -\vp \neq \rho = -\psi$ \, . }
\label{15}
\EN
The further special cases in which the inequalities in \p{15} are replaced by
equalities up to sign, {\it i.e.}
\EQ
\cases{ & $\vt = \psi = \pm \vp = \pm \rho$ \, , \cr
{\rm or} \ & $\vt= \rho = \pm \psi = \pm \vp$ \, , \cr
{\rm or} \ & $\vt= -\vp = \pm \rho = \mp \psi$\, , }
\label{151}
\EN
preserve $3/16$ supersymmetry. This includes the case discussed in
ref.~\cite{GG} in which all four angles are equal. Finally if the equal
four angles in eq.~\p{151} take the special values $\pm\frac{\pi}{2}$,
then we have 1/4 supersymmetry.

\subsection{Five angles}

For the general case of five independent angles we must return to consider
\p{7}. Supersymmetry is completely broken unless $\vt\pm \psi\pm \vp\pm \rho\pm
\zeta=0$ (mod $2\pi$). To investigate the various possibilities that this
condition allows, we set $\zeta= -\vt - \psi- \vp- \rho$ (mod $2\pi$).
Eq.~\p{7} then becomes
\bea
R^2-1 &=& 2 R \Gamma_{15} \times {\rm diag.} \left(
 \sin\vt, -\sin(\psi+\vp+\rho),\sin(\vt+\rho),
 -\sin(\psi+\vp), \sin(\vt+\vp), \right. \nn
&& -\sin(\psi+\rho), \sin(\vt+\vp+\rho), -\sin\psi,
 \sin(\vt+\psi), -\sin(\vp+\rho), \sin(\vt+\psi+\rho), \nn
&& \left.-\sin\vp, \sin(\vt+\psi+\vp), -\sin\rho,
 \sin(\vt+\psi+\vp+\rho), 0, \cdots \right),
\label{16}
\ena
showing that, generically, $1/32$ supersymmetry is preserved. However, there
are various special cases to consider when one or more of the arguments of the
sine functions vanish. Let $\rho+\psi+\vp=0$ (so $\zeta=-\vt$); one can
show that other choices give essentially the same results.
Then eq.~\p{16} becomes
\bea
R^2-1 &=& 2 R \Gamma_{15} \times {\rm diag.} \left(
 \sin\vt,0,\sin(\vt-\psi-\vp), -\sin(\psi+\vp), \right. \sin(\vt+\vp), \nn
&& \sin\vp, \sin(\vt-\psi), -\sin\psi,
 \sin(\vt+\psi), \sin\psi, \sin(\vt-\vp), \nn
&& -\sin\vp, \sin(\vt+\psi+\vp), \sin(\psi+\vp),
 \left. \sin\vt, 0, \cdots \right),
\label{17}
\ena
which yields $1/16$ supersymmetry, generically, although there are now various
subcases to consider with enhanced supersymmetry. Some of these reduce the
problem to one already considered. For example, if (in addition to the
restrictions already being considered) we set $\vp +\psi=0$ then $\rho=0$ and
the problem is reduced to the four-angle case. However, the choice
$\vp=\vt-\psi$ yields
\bea
R^2-1 &=& 2 R \Gamma_{15} \times {\rm diag.} \left(
 \sin\vt,0,0, -\sin\vt, \right. \sin(2\vt-\psi), \nn
&& \sin(\vt-\psi), \sin(\vt-\psi), -\sin\psi,
 \sin(\vt+\psi), \sin\psi, \nn
&& \sin\psi, -\sin(\vt-\psi), \sin(2\vt), \sin\vt,
 \left. \sin\vt, 0, \cdots \right),
\label{18}
\ena
which gives $3/32$ supersymmetry, a fraction not previously seen. In this case
we have $\rho=\zeta=-\vt$. There are now several subcases in which
supersymmetry
is further enhanced. These are as follows:
\begin{enumerate}
\item
$\psi=2 \vt$:  $1/8$ supersymmetry. (Here we have $\vp=\rho=\zeta=-\vt$.)\\
When $\vt=\pm \frac{\pi}{3}$ supersymmetry is further enhanced to $5/32$,
again a new fraction not previously seen.\\
When, instead, $\vt=\pm \frac{\pi}{2}$ we have $1/4$ supersymmetry.
\item
$\psi= \vt$: $3/16$ supersymmetry. (Here we have $\vp=0, \rho=\zeta=-\vt$.)\\
When $\vt=\pm \frac{\pi}{2}$ supersymmetry is again enhanced to $1/4$.
\item
$\psi=- \vt$: $1/8$ supersymmetry.
(Here we have $\vp=2\vt, \rho=\zeta=-\vt$.) \\
Again supersymmetry is enhanced to 5/32 for $\vt = \pm \frac{\pi}{3}$.\\
When, instead, $\psi=- \vt=\rho=\zeta=\pm \frac{\pi}{2},\vp=\mp \pi$,
we have $1/4$ supersymmetry.
\end{enumerate}

Other special choices do not produce anything new. For example, setting
$\vp=-\vt$ (so that $\rho=-\psi+\vt, \zeta=-\vt$) we have
\bea
R^2-1 &=& 2 R \Gamma_{15} \times {\rm diag.} \left( \sin\vt,0,
 \sin(2\vt-\psi), \sin(\vt-\psi),0,-\sin\vt,  \right.\nn
&& \sin(\vt-\psi), -\sin\psi, \sin(\vt+\psi), \sin\psi,\sin(2\vt),
 \sin\vt, \sin\psi, \nn
&& \left. \sin(\psi-\vt), \sin\vt, 0, \cdots \right)\ .
\label{20}
\ena
But this differs from \p{18} only by a permutation of the entries.
In summary the possible fractions of supersymmetry that can be
preserved by rotations with five independent angles are $1/32$, $1/16$,
$3/32$, $1/8$ and $5/32$ (which occurs only for fixed angles). The
fractions $3/16$ and $1/4$ are also possible but only if at least one of
the angles vanishes (or equals $\pm \pi$ for the latter case).

\section{Discussion}

We have found two new fractions of supersymmetry preserved by intersecting
M-brane configurations, $\nu=3/32$ and $\nu=5/32$. These fractions are
realized by particular configurations of two M-5-branes intersecting
non-orthogonally on a point. In the $5/32$ case, the relative orientation
of the two M-5-branes is completely fixed. We have arrived at these
conclusions by a purely algebraic analysis of the conditions imposed on
Killing spinors by the presence of M-5-branes with given orientations.
It would be of interest to find solutions of $D=11$ supergravity
corresponding to these new configurations. In principle, the new fractions
might also be realized on stringlike intersections of other branes, e.g.
of IIA 5-branes with 6-branes. It would be of interest to
determine whether such configurations exist.

It is known, in some cases, that there is a close relation between
the number of supersymmetries preserved by intersecting brane configurations
and reduced holonomy in Kaluza-Klein compactifications~\cite{BDL}.
This is especially clear in the case of $3/16$ supersymmetry. As shown in
\cite{GG}, and confirmed here, this fraction arises in a particular class
of intersecting M-5-brane configurations, but it also arises on
compactification of $D=11$ supergravity on (hyper-K{\"a}hler) 8-manifolds
of $Sp(2)$ holonomy. That this is not a coincidence is shown by the fact
that the $D=11$ supergravity solution corresponding to the intersecting
M-5-brane configuration is dual to a `compactification' of $D=11$
supergravity on a particular class of (non-compact) hyper-K{\"a}hler
8-manifolds \cite{GG}. In the latter case, supersymmetry is
preserved because the holonomy is reduced from $SO(8)$ to the subgroup
$Sp(2)$. In view of this it seems plausible that the new $1/16$
supersymmetric `4-angle' M-5-brane configurations found above are related
to 8-manifolds of $Spin(7)$ holonomy.
When considering `five-angle' M-5-brane configurations it
is therefore natural to wonder whether there might be a connection with
reduced holonomy subgroups of $SO(10)$. The only candidate subgroups that
are not also subgroups of $SO(8)$ are $SU(3)\times SU(2)$ and $SU(5)$,
both of which lead to $\nu=1/16$ in the context of Kaluza-Klein
compactifications of $D=11$ supergravity. There are no subgroups of
$SO(10)$ that yield the new fractions found here, $\nu=3/32, 5/32$.
Thus, while it is possible that the configurations of intersecting
M-5-branes with $\nu=1/16$ are related by duality to Kaluza-Klein
`compactifications' of $D=11$ supergravity, this is not possible for the
intersecting brane configurations preserving $3/32$ or $5/32$ supersymmetry.

The fraction $\nu=3/32$ can be obtained by compactification of the heterotic
string on hyper-K{\"a{hler} 8-manifolds because this preserves $3/16$ of the
supersymmetry of the heterotic string vacuum, which itself breaks the
supersymmetry of the M-theory vacuum. In fact, by a modification of 
arguments presented in \cite{GG}, it is not difficult to see that such
compactifications lead to $D=2$ theories with (3,0) supersymmetry. This might
seem surprising since (3,0) supersymmetric field theories are normally
automatically (4,0) supersymmetric. However, for genuine compactifications
(i.e. on compact manifolds) the lower dimensional field theory includes
gravity and there certainly do exist (3,0) $D=2$ supergravity theories. It is
of course possible to consider the same non-compact hyper-K\"{a}{hler}
8-manifolds in the context of the heterotic string as were considered
in \cite{GG} in the context of M-theory but there is no guarantee that
these field configurations will be dual to intersecting brane configurations
with $D=2$ intersections. Indeed, the above considerations based on holonomy
appear to exclude it. The $\nu=5/32$ case is rather simpler to analyse.
There are no Kaluza-Klein compactifications of any supergravity theory that
can preserve this fraction of ($D=11$) supersymmetry. Thus, the intersecting
brane interpretation is the only way in which $D=1$ $N=5$ supersymmetric or
$D=2$ (1,4) supersymmetric field theories can be obtained from M-theory.

\vskip 1cm
\noindent
{\bf Acknowledgements}: We thank George Papadopoulos, Jerome Gauntlett
and Boris Zupnik for helpful discussions. PKT acknowledges the support of
the Professor Visitante Iberdrola Program.

\newcommand{\NP}[1]{Nucl.\ Phys.\ {\bf #1}}
\newcommand{\AP}[1]{Ann.\ Phys.\ {\bf #1}}
\newcommand{\PL}[1]{Phys.\ Lett.\ {\bf #1}}
\newcommand{\CQG}[1]{Class. Quant. Gravity {\bf #1}}
\newcommand{\CMP}[1]{Comm.\ Math.\ Phys.\ {\bf #1}}
\newcommand{\PR}[1]{Phys.\ Rev.\ {\bf #1}}
\newcommand{\PRL}[1]{Phys.\ Rev.\ Lett.\ {\bf #1}}
\newcommand{\PRE}[1]{Phys.\ Rep.\ {\bf #1}}
\newcommand{\PTP}[1]{Prog.\ Theor.\ Phys.\ {\bf #1}}
\newcommand{\PTPS}[1]{Prog.\ Theor.\ Phys.\ Suppl.\ {\bf #1}}
\newcommand{\MPL}[1]{Mod.\ Phys.\ Lett.\ {\bf #1}}
\newcommand{\IJMP}[1]{Int.\ Jour.\ Mod.\ Phys.\ {\bf #1}}
\newcommand{\JP}[1]{Jour.\ Phys.\ {\bf #1}}


\begin{thebibliography}{99}
\bibitem{ZK} B.M. Zupnik and D.V. Khetselius, Yad. Fiz. {\bf 47} (1988) 1147.
\bibitem{KL} H.-C. Kao and K. Lee, Phys. Rev. {\bf D46} (1992) 4691.
\bibitem{GG} J.P. Gauntlett, G.W. Gibbons, G. Papadopoulos and P.K. Townsend,
 \NP{B500} (1997) 133, hep-th/9702202.
\bibitem{OZ} N. Ohta and J.-G. Zhou, {\it Realization of D4-branes at angles
 in super Yang-Mills theory}, hep-th/9709065.
\bibitem{HP} P.S. Howe and G. Papadopoulos, \NP{B289} (1987) 264.
\bibitem{PT1} G. Papadopoulos and P.K. Townsend, \CQG{11}
(1994) 515,  hep-th/9307066; {\it ibid.} 2163, hep-th/9406015.
\bibitem{CP1} R. Coles and G. Papadopoulos, \CQG{7} (1990) 427.
\bibitem{GPS} G.W. Gibbons, G. Papadopoulos and K.S. Stelle, {\it HKT and OKT
Geometries on Soliton Black Hole Moduli Spaces}, hep-th/9706207.
\bibitem{PT2} G. Papadopoulos and P.K. Townsend, \PL{B380} (1996)
273, hep-th/9603087.
\bibitem{TS1} A.A. Tseytlin, \NP{B475} (1996) 149, hep-th/9604035.
\bibitem{GKT} J.P. Gauntlett, D.A. Kastor and J. Traschen, \NP{B478} (1996)
 544, hep-th/9604179.
\bibitem{BDL} M. Berkooz, M.R. Douglas and R.G. Leigh, \NP{B480} (1996) 265,
 hep-th/9606139.
\bibitem{PK} P.K. Townsend, {\it M-branes at angles}, hep-th/9708074.
\end{thebibliography}
\end{document}